\documentclass[twoside]{article}

\usepackage{url}
\usepackage{epsfig}
\usepackage{graphicx}
\usepackage{amssymb,amsmath}%showkeys}
\usepackage{verbatim,enumerate}
\usepackage{verbatim,multicol,color}
\usepackage{multirow}
\usepackage{multicol}
\usepackage[english]{babel}
\usepackage{subfig}
\usepackage{color}
\usepackage{float}
\usepackage{hyperref}
\usepackage[round]{natbib}
\usepackage{setspace}
\usepackage{mathrsfs}
\usepackage{aistats2020}
\usepackage{float}
\usepackage{rotating}
\def\log{\hbox{log}}

\def\boxit#1{\vbox{\hrule\hbox{\vrule\kern6pt
          \vbox{\kern6pt#1\kern6pt}\kern6pt\vrule}\hrule}}

\def\bse{\begin{eqnarray*}}
\def\ese{\end{eqnarray*}}
\def\be{\begin{eqnarray}}
\def\ee{\end{eqnarray}}
\def\bq{\begin{equation}}
\def\eq{\end{equation}}
\def\bse{\begin{eqnarray*}}
\def\ese{\end{eqnarray*}}

\newcommand{\bI}{\mathbf{I}}

\newcommand{\bX}{\mathbf{X}}
\newcommand{\bY}{\mathbf{Y}}
\newcommand{\bZ}{\mathbf{Z}}

\newcommand{\bU}{\mathbf{U}}
\newcommand{\bV}{\mathbf{V}}

\newcommand{\bW}{\mathbf{W}}

\newcommand{\bx}{\mathbf{x}}

\newcommand{\by}{\mathbf{y}}

\newcommand{\bmu}{\boldsymbol{\mu}}

\newcommand{\0}{\mathbf{0}}

\newcommand*{\QEDB}{\hfill\ensuremath{\square}}
\DeclareRobustCommand{\rchi}{{\mathpalette\irchi\relax}}
\newcommand{\irchi}[2]{\raisebox{\depth}{$#1\chi$}} % inner command, used by \rchi

\newtheorem{thm}{Theorem}[section]

\newtheorem{lem}[thm]{Lemma}

\newcommand{\J}{\mathbf{\J}}

\begin{document}

\twocolumn[
\aistatstitle{On a Generalization of the Average Distance Classifier}
\aistatsauthor{ Sarbojit Roy  \And Soham Sarkar \And Subhajit Dutta}
\aistatsaddress{Department of Mathematics \\
and Statistics, IIT Kanpur,\\
Kanpur, India.
\And Institute of Mathematics,\\
 Ecole Polytechnique\\
 F\'{e}d\'{e}rale de Lausanne,\\
CH-1015 Lausanne, Switzerland. \And Department of Mathematics\\
and Statistics, IIT Kanpur,\\
Kanpur, India.}
]
%%%%%%%%%%%%%%%%%%%%%%%%%%%%%%%%%%%%%%%%%%%%%%%%%%%%%%%%%%%%%%%%%%%%%%%%

\begin{abstract}
In high dimension, low sample size (HDLSS) settings, the simple average distance classifier based on the Euclidean distance performs poorly if differences between the locations get masked by the scale differences. To rectify this issue, modifications to the average distance classifier was proposed by \citet*{CH2009}. However, the existing classifiers cannot discriminate when the populations differ in other aspects than locations and scales. In this article, we propose some simple transformations of the average distance classifier to tackle this issue. The resulting classifiers perform quite well even when the underlying populations have the same location and scale. The high-dimensional behavior of the proposed classifiers is studied theoretically. Numerical experiments with a variety of simulated as well as real data sets exhibit the usefulness of the proposed methodology.
\end{abstract}

\section{INTRODUCTION} \label{Intro}
Let us consider a classification problem involving two unknown multivariate distribution functions $F_1$ and $F_2$ on $\mathbb{R}^D$. Let $\rchi_1=\{\bX_{1},\ldots,\bX_{n_1}\}$ and $\rchi_2=\{\bY_{1},\ldots,\bY_{n_2}\}$ be two sets of observations from $F_1$ and $F_2$ respectively and $\rchi =  \rchi_1\cup \rchi_2$ be the training sample of size $n=n_1+n_2$. For a test point $\bZ \in \mathbb{R}^D$, our objective is to classify $\bZ$ as coming from either $F_1$ or $F_2$. Suppose that 
\begin{align*}
&T_1(\bZ)=n_1^{-1}\sum_{i=1}^{n_1}D^{-1}\|\bX_{i}-\bZ\|^2,\text{ and}\\
&T_2(\bZ)=n_2^{-1}\sum_{i=1}^{n_2}D^{-1}\|\bY_{i}-\bZ\|^2.
\end{align*}
The average distance classifier (henceforth referred to as AVG classifier) is defined as:
\begin{center}
$T(\bZ) = T_2(\bZ)-T_1(\bZ)$,\\
\end{center}
 and we classify $\bZ$ as coming from $F_1$ if $T(\bZ)>0$; otherwise we classify it as coming from $F_2$ (see \citet*{CH2009}). Here, $\|\cdot\|$ denotes the Euclidean norm on $\mathbb{R}^D$.

Let $\bmu_{jD}$ and $\Sigma_{jD}$ denote the mean vector and the dispersion matrix, respectively, corresponding to $F_j$ for $j=1,2$. Also, assume that there exist constants $\nu^2_{12}$, $\sigma^2_1$ and $\sigma^2_2$ such that 
\begin{align*}\label{const_def}
&\nu_{12}^2=\lim_{D \to \infty} \{D^{-1}\|{\bmu}_{1D}-{\bmu}_{2D}\|^2\} \mbox{ and } \\
&\sigma_j^2=\lim_{D \to \infty} \{D^{-1}{tr}(\Sigma_{jD})\} \mbox{ for } j=1,2.
\end{align*}
Here, $tr(A)$ is the sum of the diagonal elements of a $D \times D$ matrix $A$. The constants $\nu_{12}^2$ and $\sigma_1^2$, $\sigma_2^2$ are measures of the location difference and scales, respectively.
Under appropriate distributional assumptions, \citet*{HMN05} showed that if $\nu_{12}^2 \le |\sigma_1^2-\sigma_2^2|$, then the average distance classifier assigns all observations to the population having a smallest dispersion. To address this problem, \citet*{CH2009} proposed a scale adjustment to the average distance classifier, and showed that the resulting classifier performs well when $\nu^2_{12}>0$. Henceforth, we will refer to this classifier as SAVG.

Now, consider the following classification problem (say, \textbf{Example 1}) involving two $D$-dimensional Gaussian distributions $N_D(\0_D, \Sigma_D^{(1)})$ and $N_D(\0_D, \Sigma_D^{(2)})$, where $\0_D$ is the $D$-dimensional vector of zeros, and $\Sigma_D^{(1)}$ and $\Sigma_D^{(2)}$ are covariance matrices defined as follows:

\begin{center}
	$\Sigma_D^{(1)}$ =
	$
	\begin{bmatrix}
	\bI_{\lfloor\frac{D}{2}\rfloor}&&\0_{\lfloor\frac{D}{2}\rfloor \times (D-\lfloor\frac{D}{2}\rfloor)}\\
	\0_{(D-\lfloor\frac{D}{2}\rfloor)\times \lfloor\frac{D}{2}\rfloor}&&0.5\bI_{D-\lfloor\frac{D}{2}\rfloor}\\
	\end{bmatrix}
	$,\\
\vspace{0.1cm}
and\\
	\vspace{0.1cm}
$\Sigma_D^{(2)}$ = 
	$
	\begin{bmatrix}
	0.5\bI_{D-\lfloor\frac{D}{2}\rfloor}&&\0_{(D-\lfloor\frac{D}{2}\rfloor) \times \lfloor\frac{D}{2}\rfloor}\\
	\0_{\lfloor\frac{D}{2}\rfloor\times (D-\lfloor\frac{D}{2}\rfloor)}&&\bI_{\lfloor\frac{D}{2}\rfloor}\\
	\end{bmatrix}
	$.\\
\end{center}
\noindent
Here $\bI_d$ is the $d\times d$ identity matrix, $\0_{l\times m}$ is the $l\times m$ matrix of zeros and $\lfloor \cdot \rfloor$ denotes the floor function. 

\begin{figure}[H]
\centering
\captionsetup{justification=centering}
\subfloat{
\includegraphics[width= \linewidth,height=0.75\linewidth]{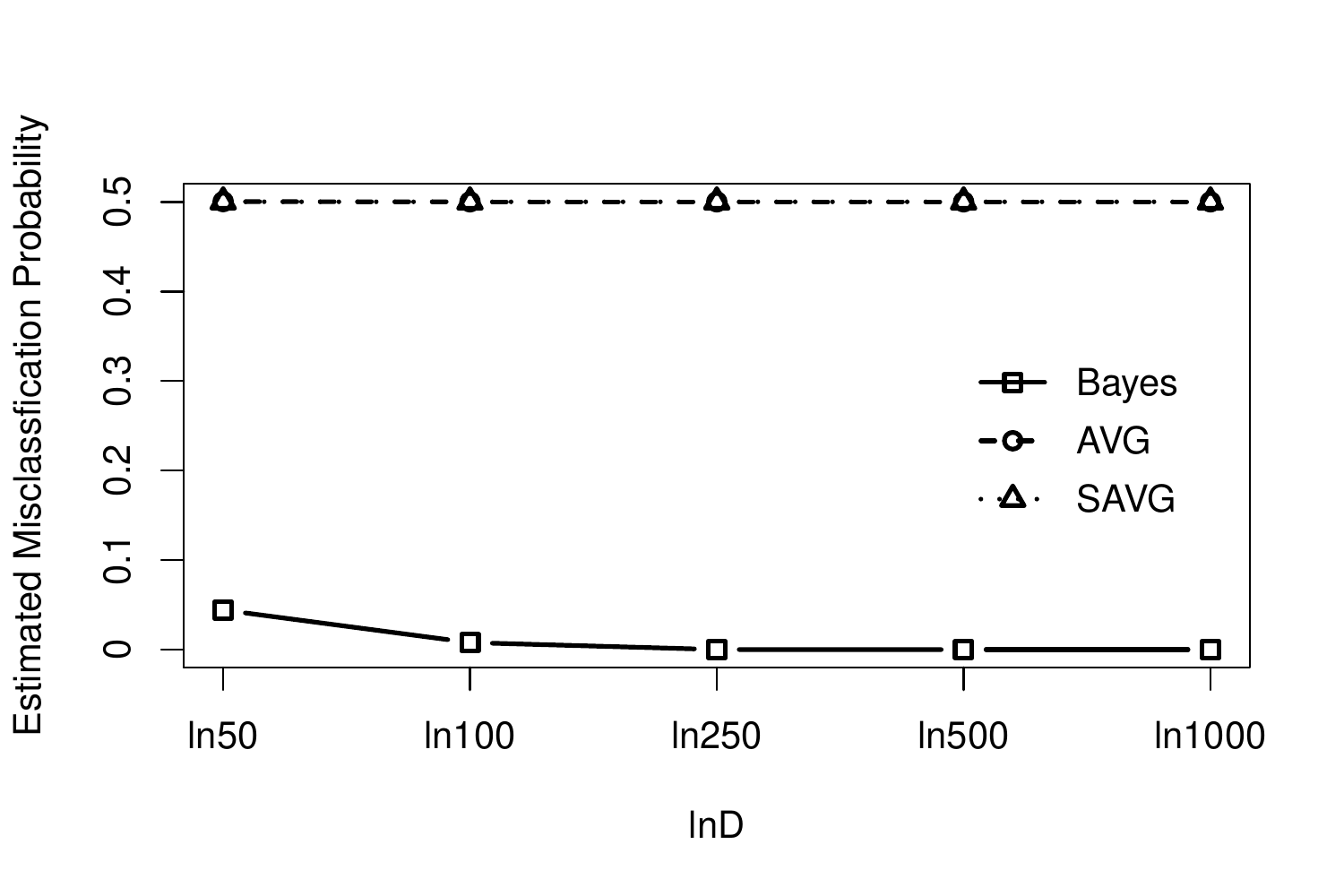}
}
\caption{ Misclassification rates of different classifiers for Example 1}
\label{plot0}
\end{figure}

We generated $50$ observations from each class to form the training sample. Misclassification rates of different classifiers are computed based on a test set consisting of $500$ ($250$ from each class) observations. This process was repeated $100$ times, and the average misclassification rates of different classifiers for varying values of $D$ are shown in Figure \ref{plot0}. Note that in this example, we have $\nu_{12}^2=0$ and $\sigma_1^2=\sigma_2^2$, i.e., the competing populations are same in terms of their location and scale parameters. As a result, though the populations differ significantly, both the classifiers performed very poorly. In fact, their performance is as good as of a classifier that puts an observation randomly to either of the classes.

\vspace{-0.35cm}
\vspace{-0.35cm}

In this article, we propose a modification of the SAVG and show that the modified classifier can discriminate between populations having different block structures. The HDLSS asymptotic properties of the proposed classifier are studied in Section \ref{general}. To implement the proposed method, we require a random vector to be partitioned into disjoint groups of components in order to extract discriminatory information from the groups. We propose some data driven methods in Section \ref{cluster} to perform the grouping. Numerical performance of the classifier on several simulated and real data sets are demonstrated in Section \ref{sim} and \ref{real}, respectively. The article ends with concluding remarks in Section \ref{conclude}. All the proofs can be found in the supplementary material.
\section{GENERALIZED AVERAGE DISTANCE CLASSIFIER} \label{general}
The average distance classifier (AVG) assigns a new observation $\bZ$ to $F_1$ if $T(\bZ)>0$; otherwise to $F_2$. Note that $D\ E[T(\bZ)]=\pm\|\bmu_{1D}-\bmu_{2D}\|^2 + tr(\Sigma_{2D}-\Sigma_{1D})$ according as $\bZ\sim F_1$ or $F_2$.
%$E[T(\bZ)]=\|\bmu_{1D}-\bmu_{2D}\|^2 + tr(\Sigma_{2D}-\Sigma_{1D})$ if $\bZ\sim F_1$, and $E[T(\bZ)]=-\|\bmu_{1D}-\bmu_{2D}\|^2+ tr(\Sigma_{2D}-\Sigma_{1D})$ when $\bZ\sim F_2$.
As $D\to\infty$, $\bZ$ is correctly classified if $\nu_{12}^2>|\sigma^2_1-\sigma^2_2|$. So, the classifier is useful when the location difference is larger than the scale difference.
\citet*{CH2009} identified that the scale factor $tr(\Sigma_{2D}-\Sigma_{1D})$ is a nuisance parameter, and made the following adjustment to the average distance classifier:
\begin{align*}
&T^{adj}_1(\bZ)=n_1^{-1}\sum_{i=1}^{n_1}D^{-1}\|\bX_{i}-\bZ\|^2-D(\rchi_1|\rchi_1)/2\text{ and}\\
&T^{adj}_2(\bZ)=n_2^{-1}\sum_{i=1}^{n_2}D^{-1}\|\bY_{i}-\bZ\|^2-D(\rchi_2|\rchi_2)/2,\\
&\text{where }\mathcal{D}(\rchi_1|\rchi_1)=\{n_1(n_1-1)\}^{-1}\hspace{-0.35cm}\sum_{1\leq l\neq k\leq n_1}\hspace{-0.35cm}D^{-1}\|\bX_{l}-\bX_{k}\|^2,\\
&\mathcal{D}(\rchi_2|\rchi_2)=\{n_2(n_2-1)\}^{-1}\hspace{-0.35cm}\sum_{1\leq l\neq k\leq n_2}\hspace{-0.35cm}D^{-1}\|\bY_{l}-\bY_{k}\|^2,\text{ and}\\
&\text{ proposed the classifier }T^{adj}(\bZ)=T^{adj}_2(\bZ)-T^{adj}_1(\bZ).
\vspace{-0.2cm}
\end{align*}
Observe that $E[D(\rchi_{j}|\rchi_j)]=2tr(\Sigma_{jD}),~ j=1,2$. Thus,  $ D\ E[T^{adj}(\bZ)]=\pm\|\bmu_{1D}-\bmu_{2D}\|^2$ according as $\bZ\sim F_1$ or $F_2$, and it perfectly classifies a new observation whenever $\nu_{12}^2>0$ as $D\to \infty$.

In \textbf{Example  1}, we observed that both the AVG and the SAVG classifiers exhibit poor results (see Figure \ref{plot0}) since $F_1$ and $F_2$ do not differ in terms of location and scale parameters, i.e., $\sigma^2_1=\sigma^2_2$ and $\nu_{12}^2=0$. 

To circumvent this problem, we consider a new class of dissimilarities.  Let $\bU^{\top}=(\bU^{\top}_1,\ldots,\bU^{\top}_B)$ represent of $D$-dimensional random vector $\bU$, where $\bU_b\in\mathbb{R}^{D_b}$, $1\leq D_b<D_0<\infty$ with $\sum_{b=1}^{B}D_b=D$. For two vectors $\bU$ and $\bV$, we define the dissimilarity as 
% \vspace{-0.4cm}
\begin{equation}\label{betadef}
 \vspace{-0.1cm}
 h^B_{\gamma}(\bU,\bV)=B^{-1}\sum_{b=1}^B\gamma(D^{-1}_b\|\bU_b - \bV_b\|^2).
\end{equation}
Here, $\gamma:\mathbb{R}^+ \to \mathbb{R}^+$ is a continuous, monotonically increasing function with $\gamma(0)=0$. We call $\bU_b$ the $b$-th block of $\bU$. As collections of component variables, $\bU_b$ and $\bU_{b^{\prime}}$ are disjoint for $b\neq b^{\prime}$. Therefore, $\{\bU_1,\ldots,\bU_B\}$ constitutes a partition of $\bU$.

We use this class of dissimilarities to modify the SAVG classifier. For a test point $\bZ \in \mathbb{R}^D$, we define 
\begin{align*}
&T^{\gamma}_1(\bZ)=n^{-1}_1\sum_{i=1}^{n_1}h^B_{\gamma}(\bZ,\bX_{i})-D_{\gamma}(\rchi_{1}|\rchi_1)/2\text{ and}\\
&T^{\gamma}_2(\bZ)=n^{-1}_2\sum_{i=1}^{n_2}h^B_{\gamma}(\bZ,\bY_{i})-D_{\gamma}(\rchi_{2}|\rchi_2)/2,\\
&\text{where}\\
&D_{\gamma}(\rchi_{1}|\rchi_1)=\{n_1(n_1-1)\}^{-1}\hspace{-0.3cm}\sum_{1\leq l\neq k\leq n_1}\hspace{-0.3cm}h^B_{\gamma}(\bX_{l},\bX_{k})\text{ and}\\
&D_{\gamma}(\rchi_{2}|\rchi_2)=\{n_2(n_2-1)\}^{-1}\sum_{1\leq l\neq k\leq n_2}\hspace{-0.3cm}h^B_{\gamma}(\bY_{l},\bY_{k}).
\end{align*}
Here, $D_{\gamma}(\rchi_{j}|\rchi_j)$ is the within class deviation of $F_j$ in terms of the new dissimilarity $h^B_{\gamma}$ for $j=1,2$. The proposed average distance classifier is $T^{\gamma}(\bZ)=T^{\gamma}_2(\bZ) - T^{\gamma}_1(\bZ)$ and we classify $\bZ$ as coming from $F_1$ if $T^{\gamma}(\bZ)>0$, otherwise $\bZ$ is classified as coming from $F_2$.

The dissimilarity index proposed in \eqref{betadef} has certain advantages as we will see in the next subsection. For the time being, observe that if $\gamma(t)=t,~t\geq 0$ and $B=D$, then $h^B_{\gamma}$ reduces to the squared Euclidean distance scaled by $D$. Consequently, the proposed classifier coincides with the SAVG classifier. Therefore, the new classifier can be viewed as a generalization of the later. We call it generalized Scale Adjusted Average Distance (gSAVG) Classifier. In the next subsection, we study the behavior of gSAVG in the HDLSS asymptotic regime, where the sample size $n$ is assumed to be fixed and the dimension $D$ increases to infinity.

\subsection{Behavior of gSAVG in the HDLSS Asymptotic Regime} \label{HDLSS}
Let $F^b_j$ be the distribution of $\bU_b$ when $\bU\sim F_j,~1\leq b\leq B,~ j=1,2$. 
To study the asymptotic behavior of the gSAVG classifier we make the following assumptions: 
\begin{align*}
(A1) &~E[\gamma^2(D_b^{-1}\|\bU_b-\bV_b\|^2)]\leq C_1<\infty\ \forall\ 1\leq b\leq B. \\%\label{ass4} 
(A2) &\hspace{-0.4cm}\sum_{1\leq b<b^\prime\leq B}\hspace{-0.4cm}Corr(\gamma(D_b^{-1}\|\bU_b-\bV_b\|^2),\gamma(D_{b^\prime}^{-1}\|\bU_{b^\prime}-\bV_{b^\prime}\|^2))\\
&=o(B^2).%\label{ass5}
\end{align*}
\vspace{-0.5cm}

It is evident that $(A1)$ is satisfied if $\gamma$ is bounded.
% However, if we choose $\gamma$ to be bounded on $\mathbb{R}^+$, then the assumption of boundedness of the components can be relaxed.
Assumption $(A2)$ holds if the groups of component variables of the underlying populations are independent. However, it holds even when the groups are dependent with some additional conditions on their dependence structure. For instance, in the case of sequence data, it holds when the sequence has the $\rho$-mixing property (see, e.g. \citet*{HMN05},  \citet*{bradley2005basic}). Under assumptions $(A1)$ and $(A2)$, the high-dimensional behavior of the dissimilarity index $h^B_{\gamma}$ and the gSAVG classifier are given by the following lemma.

\begin{lem}\label{secondlemma}
Suppose that $\bU\sim F_j$ and $\bV\sim F_{j^\prime}$ with $ j,j^{\prime} \in \{1,2\}$ are two independent random vectors satisfying $(A1)$ and $(A2)$. Then $$\big|h^B_{\gamma}(\bU,\bV)-\tilde{h}^B_{\gamma}(j,j^\prime)\big|\stackrel{P}{\to}0\text{ as }B\to\infty ,$$ where $\tilde{h}^B_{\gamma}(j,j^\prime)=B^{-1}\sum_{b=1}^BE\{\gamma(D_b^{-1}\|\bU_b - \bV_b\|^2)\}$. Consequently, for a test observation $\bZ$, independent of $\rchi$, we have 
$$\big|T^{\gamma}(\bZ)-E[T^{\gamma}(\bZ)]\big|\stackrel{P}{\to}0\ as\ B\to\infty.$$
\end{lem}

Observe that
$$
E[T^{\gamma}(\bZ)]=
\begin{cases}
\psi^{\gamma}_B(1,2)/2,\text{ if }\bZ\sim F_1,\text{ and}\\
-\psi^{\gamma}_B(1,2)/2, \text{ if } \bZ\sim F_2.
\end{cases}
$$
Here, $\psi^{\gamma}_B(1,2)=2\tilde{h}^B_{\gamma}(1,2)-\tilde{h}^B_{\gamma}(1,1)-\tilde{h}^B_{\gamma}(2,2)$. Let us have a closer look at the quantity $\psi^{\gamma}_B(1,2)$. It can be expressed as the following:
\begin{align*}
&\psi^{\gamma}_B(1,2) =B^{-1}\sum_{b=1}^Be_{\gamma}(F^b_1, F^b_2),\text{ where }\forall 1\leq b\leq B,\\
&e_{\gamma}(F^b_1,F^b_2)=E\big[2{\gamma(D_b^{-1}\|\bX_{1b}-\bY_{1b}\|^2)}\\
&-{\gamma(D_b^{-1}\|\bX_{1b}-\bX_{2b}\|^2)}
-{\gamma(D_b^{-1}\|\bY_{1b}-\bY_{2b}\|^2)}\big] .
\end{align*}
This quantity $e_{\gamma}(F^b_1,F^b_2)$ is an energy distance between $F^b_1$ and $F^b_2$ and is non-negative. \citet*{BF10} showed that for appropriate choices of $\gamma$, $e_{\gamma}(F,G)\geq 0$, and the equality holds if and only if $F = G$ for any $F$ and $G$. The following lemma states suitable choices of $\gamma$ in this context.

\begin{lem}\label{sixthlemma}
If $\gamma$ has non-constant completely monotone derivative on $\mathbb{R}^+$, then $\psi^{\gamma}_B(1,2)=0$ if and only if $F^b_1=F^b_2$ for all $1\leq b\leq B$.
\end{lem}

\noindent
The quantity $\psi^{\gamma}_B(1,2)$ is nothing but the simple average of the energy distances $e_{\gamma}(F^b_1, F^b_2), 1\leq b\leq B$. Hence, it can be interpreted as an average energy distance between $F_1$ and $F_2$. Lemma \ref{sixthlemma} ensures that there will be a positive energy distance between $F_1$ and $F_2$ as long as the distributions of the blocks are different. This property of $\psi^{\gamma}_B$ is useful in discriminating two populations since it works as a measure of separation between them. It is reasonable to assume the following:
\begin{align*}
(A3)~\liminf_{B\to\infty}\psi^{\gamma}_B(1,2)>0\hspace{3in}
\end{align*}
This assumption ensures that the separation between the two populations is asymptotically not negligible. When combined with Lemmas \ref{secondlemma} and \ref{sixthlemma}, this also ensures that for appropriate choice of $\gamma$, as $D\to\infty$, the classifier $T^{\gamma}(\bZ)$ converges in probability to a strictly positive value (respectively, strictly negative value) if $\bZ\sim F_1$ (respectively, $\bZ\sim F_2$). This brings us to the following theorem which states the behavior of the proposed classifier in the HDLSS asymptotic regime.
\begin{thm}\label{thm0}
If $\gamma:\mathbb{R}^+ \to \mathbb{R}^+$ is a continuous, monotonically increasing function having non-constant completely monotone derivative such that $\gamma(0)=0$ and assumptions \textrm{(A1)-(A3)} are satisfied, then the misclassification probability of the gSAVG classifier converges to zero as $D\to \infty.$
\end{thm}

There are several choices of $\gamma$ that satisfy the conditions stated in Theorem \ref{thm0} (see \citet*t*[p.1338]{BF10}) e.g., namely, $\gamma_1(t)=1-e^{-t}$, $\gamma_2(t)=\sqrt{t}/2$ and $\gamma_3(t)=\log(1+t)$, etc.

Recall that the misclassification probability of the gSAVG classifier depends on the blocks, or the partition of vector $\bU$, i.e., $\{\bU_1,\ldots, \bU_B\}$ (see Definition \ref{betadef}). In practice, these blocks are unknown; hence need to be estimated. The following section exhibits data driven methods of estimating these clusters.

\section{ESTIMATION OF BLOCKS}\label{cluster}
Our goal is to partition a $D$-dimensional random vector $\bU=(U_1,\ldots,U_D)^{\top}$ into $B$ disjoint blocks. This is essentially a problem of variable clustering. We assume that if $U_d$ is a member of the $b$-th block for $F_1$, then $U_d$ is a member of the $b$-th block for $F_2$ as well, for $1 \leq b \leq B$ (i.e., the number of groups and its members are identical in both populations). Under this assumption, each variable has $n~(=n_1+n_2)$ observations. One can view this as a problem in clustering of $D$ data points in $\mathbb{R}^n$. Any appropriate clustering method (see, e.g., \citet*{friedman2001elements}) can be used for this purpose. 

Here, we use the hierarchical clustering method with the average linkage to form the blocks. In the context of our problem, a meaningful way to form the blocks is to keep the components together that have strong pairwise correlations between them. The sign of correlation is not important in this case. Be it positive or negative, we consider two components to be similar if they have strong correlation, and components having weak correlations are put into different clusters. We propose to use $l(d,d^{\prime})=1-|\rho(d,d^{\prime})|$ as the measure of  dissimilarity while defining the linkage for hierarchical clustering. Here, $\rho(d,d^\prime)$ is the sample correlation coefficient between $\bW_d$ and $\bW_{d^\prime}$, where $\bW_d = (X_{1d},X_{2d},\ldots,X_{n_1d},Y_{1d},Y_{2d},\ldots,Y_{n_2d}), 1\leq d\leq D$.
For heavy-tailed distributions, a robust measure of correlations can be used for this purpose. There are several methods available in the literature for estimating high-dimensional robust correlation matrix (see \citet*{raymaekers2017fast}).

A smaller value of $l(d,d^\prime)$ indicates that the $d$-th and the $d^{\prime}$-th components should be put together in the same cluster. Once we have the pair-wise dissimilarities for all pairs, we implement hierarchical clustering on the set of components.
Each stage in the hierarchy induces a set of blocks, and the whole hierarchy represents a nested structure among the blocks obtained at different levels of dependence. The dendrogram associated with hierarchical clustering provides a interpretable visual summary of the blockings. Now, one needs to decide an appropriate cutoff point in the dendrogram to yield meaningful disjoint blocks. Suppose that $H$ is the set of all heights that are obtained at each step for merging two clusters. We order the values in $H$, and find the $p$-th percentile for different values of $p \in [0,1]$. In our numerical work, we have discretized this set as $P=\{0,0.1,\ldots,0.9,1\}$. Let $h_p$ denote the $p$-th percentile. For each fixed $p$, we cut the dendrogram at $h_p$ and obtain a set of clustered components. As $p$ increases, the number of clusters decreases, while the size of each cluster increases. In other words, $h_0$ corresponds to the scenario where each cluster consists of one component variable only, i.e., $B=D$. On the other hand, $h_1$ leads to the setting where there is only one cluster consisting of all the $D$ component variables. 

It is clear that construction of the `optimal' blocking depends crucially on the choice of $p$. Our aim is to select the blocking that yields the minimum misclassification rate. To achieve this, we use the leave-one-out cross-validation method (see, e.g., \citet*{friedman2001elements}). For a fixed value of $p \in P$, define 
$$e_p=\sum_{\bU\in\rchi} \mathbb{I}\{\text{gSAVG}^{-\bU}(\bU) \neq\text{ true label of }\bU\}/ (n_1+n_2).$$ 
Here, the classifier gSAVG$^{-\bU}$ is constructed by leaving out the observation $\bU$ from the training data $\rchi$. Define $\hat p=\arg\min_{p\in P}{e_p}$. For the observations of the test data, we consider the clusters induced by $h_{\hat p}$, and carry out further analysis.

\section{SIMULATION STUDIES} \label{sim}

\begin{table*}[!htbp]
	\begin{center}
	{\scriptsize
		\centering
		\caption{ Misclassification rates and standard errors (stated within brackets) of different classifiers for $D=1000$.}
		%\hspace*{0.25in}
		\renewcommand{\arraystretch}{1.2}
		\begin{tabular}{|c|c|c|c|c|c|c|c|c|}
		\hline
		Example $\downarrow$ & GLMNET & RF & NN-RAND & SVM-LIN & SVM-RBF & AVG & SAVG & gSAVG \\ \hline
			
%		\begin{tabular}{|@{\hskip1pt}c@{\hskip1pt}|@{\hskip1pt}c@{\hskip1pt}|@{\hskip1pt}c@{\hskip1pt}|@{\hskip1pt}c@{\hskip1pt}|@{\hskip1pt}c@{\hskip1pt}|@{\hskip1pt}c@{\hskip1pt}|@{\hskip1pt}c@{\hskip1pt}|@{\hskip1pt}c@{\hskip1pt}|@{\hskip1pt}c@{\hskip1pt}|}
%			\hline
%			\multicolumn{1}{|@{\hskip1pt}c@{\hskip1pt}}{Ex} & \multicolumn{1}{|@{\hskip1pt}c@{\hskip1pt}}{GLMNET}  & \multicolumn{1}{|@{\hskip1pt}c@{\hskip1pt}}{RF} &  \multicolumn{1}{|@{\hskip1pt}c@{\hskip1pt}}{NN-RAND}  & \multicolumn{1}{|@{\hskip1pt}c@{\hskip1pt}}{SVM-LIN} & \multicolumn{1}{|@{\hskip1pt}c@{\hskip1pt}}{SVM-RBF}& \multicolumn{1}{|@{\hskip1pt}c@{\hskip1pt}}{AVG} &  \multicolumn{1}{|@{\hskip1pt}c@{\hskip1pt}}{SAVG} & \multicolumn{1}{|@{\hskip1pt}c@{\hskip1pt}}{gSAVG}\\
%			\hline
			1     & 0.4677  & 0.0128  & 0.3977 & 0.4973  & 0.4844  & 0.5000  & 0.5000  & {\bf 0.0000} \\
			& (0.0184) & (0.0059) & (0.0245) & (0.0240) & (0.0235) & (0.0000) & (0.0000) & (0.0000)\\
			\hline
			2     & 0.4736 & 0.4827 & 0.4114 & 0.5001 & 0.4893 & 0.4994 & 0.5000 &  {\bf 0.0123}\\ %gamma1
			& (0.0175)  & (0.0243) & (0.0233) & (0.0197) & (0.0183) & (0.0162) & (0.0000) & (0.0067)\\
			\hline
			3     & 0.4746 & 0.3454 & 0.4636 & 0.5009 & 0.4952 & 0.4999 & 0.5000 & {\bf 0.2536} \\ %gamma2
			& (0.0183) & (0.0248) & (0.0211) & (0.0197) & (0.0207) & (0.0004) & (0.0007) & (0.0198)\\
			\hline	
		\end{tabular}%
		\label{simtable}%
	}
	\end{center}
\end{table*}%

In this section, we analyze some high-dimensional simulated data sets to compare the performance of gSAVG classifier with the AVG classifier and the SAVG classifier. Along with \textbf{Example 1} introduced in Section \ref{Intro}, we consider two more examples in this section for this purpose.

Suppose that $\{X_d: d\in \mathbb{N}\}$ and $\{Y_d: d\in \mathbb{N}\}$ are two sequences of random variables such that $X_d \sim N(0,1)$, $X_d\stackrel{L}{=}Y_d$, and $X_d\perp Y_d \text{ for all }d\geq 1$. Let $\bX$ and $\bY$ be two $D$-dimensional random vectors such that 
$\bX=(X_1,X_2, sign(X_4)X_3, sign(X_3) X_4, X_5, X_6, \newline sign(X_8) X_7, sign(X_7) X_8,\ldots)^T$ and $\bY=(sign(Y_2) Y_1, \newline sign(Y_1) Y_2,Y_3,Y_4,sign(Y_6) Y_5, sign(Y_5) Y_6,\ldots)^T$, where $sign(u) = \pm 1$ if $u \gtrless 0$.
For \textbf{Example 2}, we consider $F_1$ and $F_2$ to be the distributions of $\bX$ and $\bY$, respectively. Observe that the one dimensional marginals are all $N(0,1)$ for both $F_1$ and $F_2$. Hence, $\nu^2_{12}=0$ and $\sigma^2_1=\sigma^2_2$. However, the dependence structures of $F_1$ and $F_2$ are different. We construct \textbf{Example 3} in similar way, but in this case $\{X_d:  d\in \mathbb{N}\}$ and $\{Y_d: d\in \mathbb{N}\}$ are sequences  of independent and identically distributed random variables following univariate Cauchy distribution with location 0 and scale 1.

In each example, we simulated data for $D=50$, $100$, $250$, $500$ and $1000$. The training sample was formed by generating $50$ observations from each of the two classes, while a test set of size $500$ ($250$ from each class) was used. This process was repeated $100$ times to compute the estimated  misclassification rates of the different classifiers, which are reported in Table \ref{simtable} along with their corresponding standard errors.

Recall that we have the functions $\gamma_1$, $\gamma_2$ and $\gamma_3$ as possible choices for assessing the performance of the gSAVG classifier. The numerical results for \textbf{Examples 1-3} are reported in Table \ref{simtable}. In {\bf Example 2} and {\bf 3}, we  observed that $\gamma_1$ performed better than the others. So we have reported the results based on it only (see Figure \ref{plot1} and Table \ref{simtable}). For \textbf{Example 3}, we used a bounded $\gamma$ function to ensure that assumptions $(A1)$ and $(A2)$ hold. So, we have reported the misclassification rate for $\gamma_1$ only (see the Table %\ref{table.sim.Appendix} 
in the supplementary material for other choices of $\gamma$).  As we can observe that the minimum misclassification rate was attained by gSAVG (also see the Table in the supplementary material
% \ref{table.sim.Appendix} 
for the complete result). It is evident from Figure \ref{plot1} that the usual average distance classifier and the scale adjusted version of it performed poorly in all three examples. However, gSAVG outperformed all the classifiers and led to perfect classification in high dimensions for {\bf Example 1} and {\bf 2}. In \textbf{Example 3}, it continued to perform way better than the others showing a steady decrease in misclassification rate with increasing dimension of the data $D$. It misclassifies 25\% of the test observations at $D=1000$. However, when the blocking is known, the misclassification rate goes down to zero. Recall that in {\bf Example 3} we estimate the blocks based on robust estimate of scale matrix as described in Section \ref{cluster}. The deterioration in performance of gSAVG is due to the error in estimation, better estimates may lead to improved classification.

We have also compared the performance of the gSAVG in high dimension with some well-known classifiers. The training and test sets remain the same as before with sizes $50$ ($25+25$) and $500$ ($250+250$), respectively. This procedure was iterated $100$ times. The average misclassification rates along with the corresponding standard errors are reported in Table \ref{simtable}. Performance of GLMNET (see \citet*{friedman2001elements}), random forest (referred to as RF) (see \citet*{breiman2001random}) and NN classifiers based on the random projection method (referred to as NN-RAND) (see \citet*{deegalla2006reducing}) were studied.
\begin{figure}[H]
%\vspace{-0.25cm}
\centering
\captionsetup{justification=centering}

\subfloat{
\includegraphics[width= \linewidth,height=0.75\linewidth]{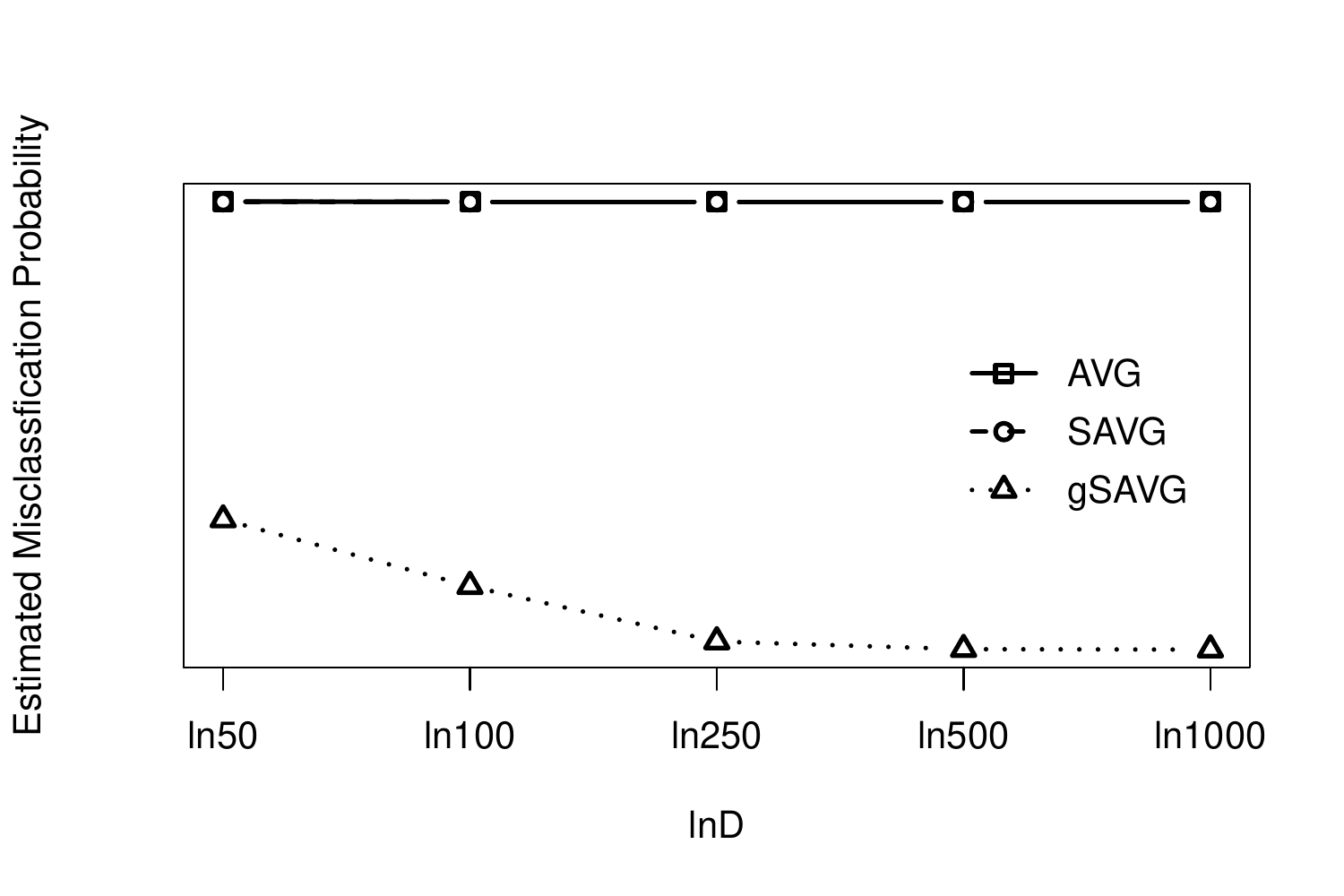}
}
%\end{figure}
\vspace{-0.75cm}
%\begin{figure}[H]
\subfloat{
\includegraphics[width= \linewidth,height=0.75\linewidth]{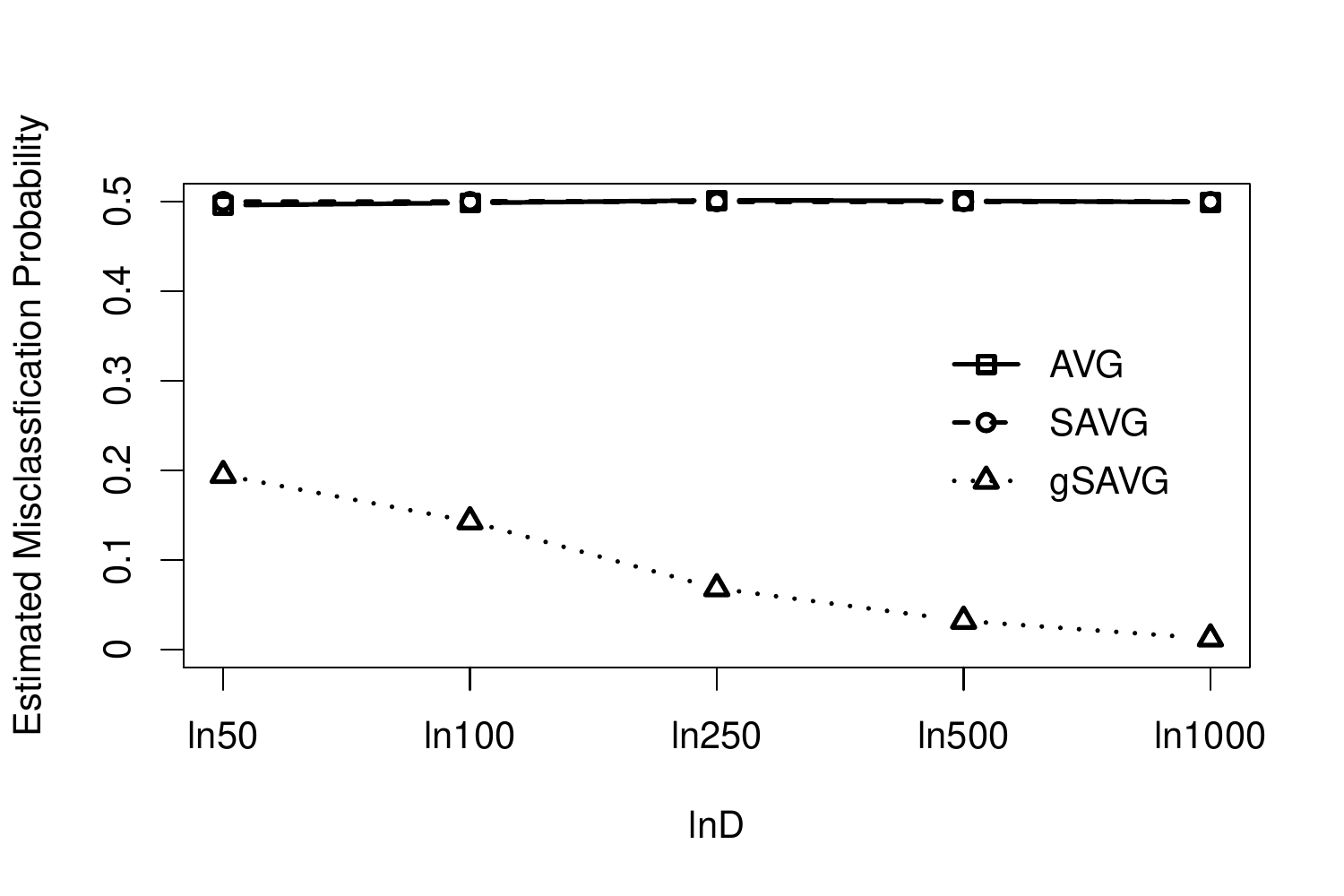}
}
\end{figure}

\begin{figure}[H]
\ContinuedFloat
\vspace{-0.75cm}
\subfloat{
\includegraphics[width= \linewidth,height=0.75\linewidth]{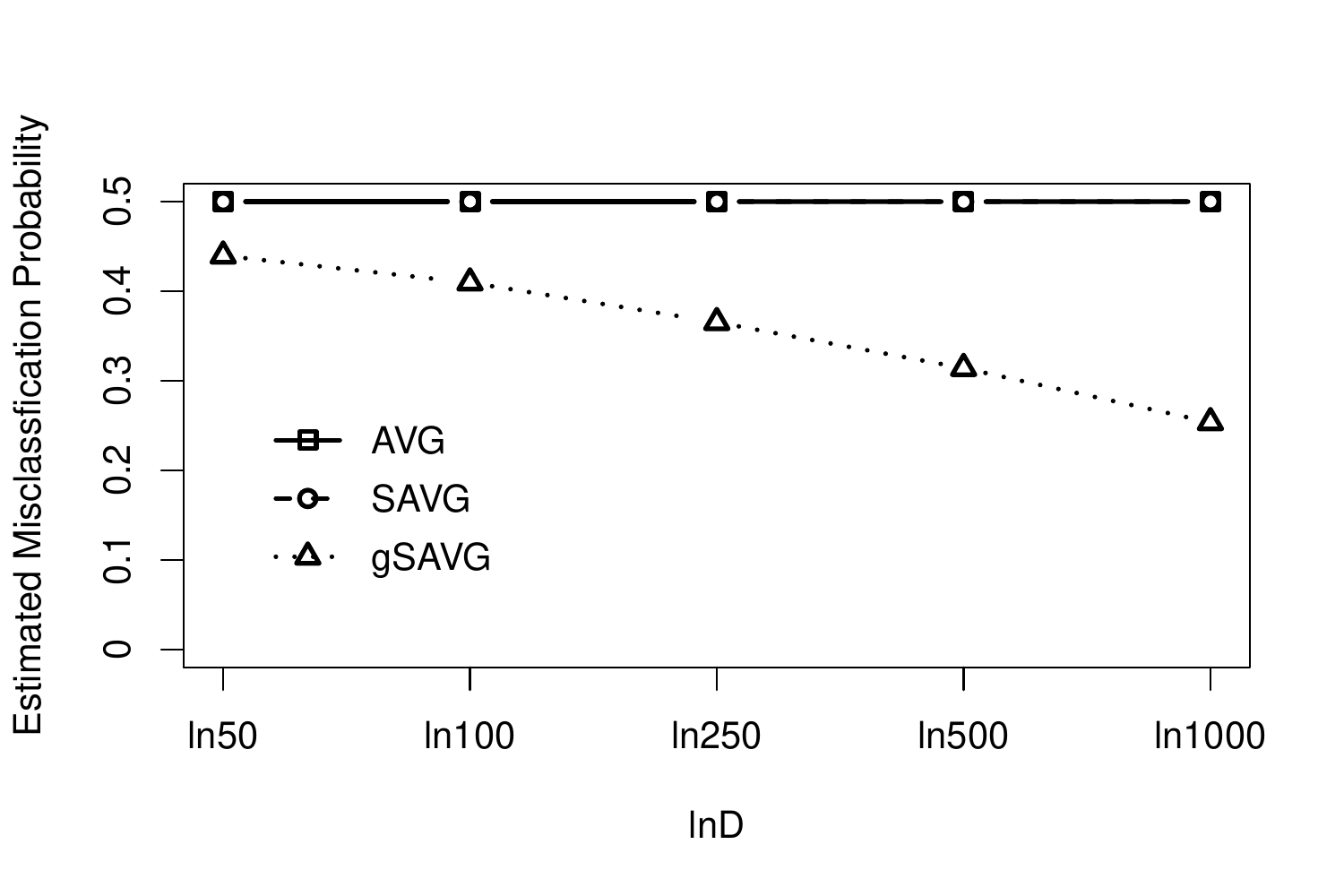}
}
\caption{ Misclassification rates of different classifiers for Example 1 (top), 2 (middle) and 3 (bottom)}
\label{plot1}
\end{figure} 
Misclassification rates of the linear and non-linear support vector machines (SVM) are reported as well. For non-linear SVM, we used the radial basis function (RBF) kernel, i.e., $K_\theta(\bx,\by)=\mathrm{exp}\{-\theta\|\bx-\by\|^2\}$ (see \citet*{vapnik1998statistical}) with the default value of the regularization parameter $\theta=1/D$. We used the {\tt R} package {\tt glmnet} for implementation of GLMNET. The {\tt randomForest}, {\tt RandPro} and {\tt e1071} packages were used for RF, NN-RAND and SVM, respectively.

To summarize Table \ref{simtable}, we observe that our proposed gSAVG classifier outperformed all the other classifiers for all examples. In fact, most of the classifiers yielded a high misclassification rate with some of them having nearly $50$\% misclassification (equivalent to the outcome of random classification). On the other hand, advantage of using the generalized dissimilarity index is clear from the superior performance of gSAVG in these examples. Interestingly, RF yielded a competitive misclassification rate in \textbf{Example 1}. The underlying class boundary is quadratic here, and RF was probably extracting information from some low-dimensional partitions at $D=1000$ and correctly classified most of the test data points (also see \citet*{FCB_2014} for more discussions).

\section{REAL DATA ANALYSIS} \label{real}
We further analyzed some benchmark data sets for assessment of our proposed gSAVG classifiers. The {\tt HouseTwenty} and {\tt Computer} data are available at the UCR Time Series Archive (2018) (see \citet*{UCRArchive2018}). {\tt GSE3726} and {\tt Leukemia} data sets are taken from \url{http://www.biolab.si/supp/bi-cancer/projections/}.
% has $905$ data points across $50$ classes with unequal class proportions. The data points have dimension $270$. The data sets {\tt CricketX, CricketY} and {\tt CricketZ} have 780 data points distributed over 12 classes. Each class has 65 data points and the data are of dimension 300. The {\tt HouseTwenty} data set has 2000-dimensional observations from $2$ classes with $89$ observations from the first and $70$ observations from the second class. The {\tt PigCVP} dataset consists of a $312$ observations of dimension $2000$. This data set poses a $52$ class classification problem. The samples are distributed equally among the classes only in the series of {\tt Cricket} data sets.
Detailed descriptions of all these data sets are also available at the sources.

\begin{table}[htbp]
\begin{center}
{
%\vspace{-0.1in}
\scriptsize
\centering
\caption{Description of benchmark data sets}
%\hspace*{1.5cm}

\renewcommand{\arraystretch}{1.25}
\begin{tabular}{|c|c|c|c|c|c|c|} \hline
%\begin{tabular}{|@{\hskip1pt}c@{\hskip1pt}|@{\hskip1pt}c@{\hskip1pt}|@{\hskip1pt}c@{\hskip1pt}|@{\hskip1pt}c@{\hskip1pt}|@{\hskip1pt}c@{\hskip1pt}|@{\hskip1pt}c@{\hskip1pt}|@{\hskip1pt}c@{\hskip1pt}|}
%\hline
%\multicolumn{1}{|@{\hskip1pt}c@{\hskip1pt}|}{Data $\to$} & \multicolumn{1}{@{\hskip1pt}c@{\hskip1pt}|}{GSE3726} & \multicolumn{1}{@{\hskip1pt}c@{\hskip1pt}|}{HouseTwenty} & \multicolumn{1}{@{\hskip1pt}c@{\hskip1pt}|}{Leukemia} & \multicolumn{1}{@{\hskip1pt}c@{\hskip1pt}|}{Prostate} \\
{Data $\to$} & Computer & {GSE3726} & {HouseTwenty} & {Leukemia} \\
\hline
$J$ & $2$ & $2$ & $2$ & $2$ \\ %gamma2
\hline
$D$ & $720$ & $22283$ & $2000$ & $5147$ \\ %gamma1
\hline
$n$ & $500$ & $52$ & $159$ & $72$\\    %gamma1
\hline
$n_1, n_2$ & $250, 250$ & $21, 31$ & $70, 89$ & $47, 65$ \\    %gamma1
%\hline
%Test  & $$ &  &  &  &  & \\    %gamma1
\hline
%$n_i$ &  &  &  &  &  & \\    %gamma1
%\hline
\end{tabular}%
\label{realdata}%
}
\end{center}
\end{table}%

\noindent
For a data set, we randomly selected $50$\% of the observations (without replacement) corresponding to each class to form a training set. The rest of the observations were considered as test cases. This procedure was repeated $100$ times over different splits of the data to obtain more stable estimates of the misclassification probabilities.
%to compute the estimated  Misclassification rates of the different classifiers, which are reported in Table \ref{realdata} along with their corresponding standard errors.
%The  Misclassification rates of the NN classifiers based of the Euclidean distance, MADD, gMADD and ggMADD are estimated based on these newly constructed training and test data sets.  Misclassification rates of SVM-LIN, SVM-RBF, GLMNET, NN-RAND and RF are also computed. We repeat this procedure $100$ times. 
The mean and standard deviation (stated witin bracket) of these $100$ estimates %for the misclassification rate corresponding to the different classifiers 
are reported in Table \ref{realdata}. 
It is to be noted that for gSAVG classifier, we carried out the analysis for all three choices of the function $\gamma$ (as described in Section \ref{general}). However, we report only the minimum misclassification rate for each data set. %For {\tt GSE3726}, the function $\gamma_2$ yielded the minimum misclassification rate of gSAVG, whereas $\gamma_3$ lead to the lowest misclassification rate for the {\tt HouseTwenty} data. For the other data sets, $\gamma_1$ was the one that yields the lowest misclassification rates for the NN-ggMADD classifier.
%The detailed results are provided in the appendix. %supplementary material.

\begin{table*}[htbp]
\begin{center}
{
\caption{Misclassification rates and standard errors (stated within brackets) of different classifiers in benchmark data sets}
\renewcommand{\arraystretch}{1.2}
\scriptsize
\begin{tabular}{|c|c|c|c|c|c|c|c|c|}
\hline
Data Sets $\downarrow$ & GLMNET & RF & NN-RAND & SVM-LIN & SVM-RBF & AVG & SAVG & gSAVG \\ \hline
Computer & 0.3910 & 0.3833 & 0.4217 & 0.4616 & 0.3982 & 0.4982 & 0.4796 & {\bf 0.3611}\\
& (0.0244) & (0.0242) & (0.0275) & (0.0315) & (0.0265) & (0.0222) & (0.0217) & (0.0192)\\ \hline
GSE3726 &  0.1339 & 0.1345 & 0.1231 & 0.1035 & 0.3751 &  0.3385 & 0.2873 & \textbf{0.0946} \\ %gamma4
& (0.0178) & (0.0162) & (0.0492) & (0.0194) & (0.0150) & (0.0975) & (0.0996) & (0.0732) \\ \hline
HouseTwenty   & 0.2470 & 0.1466 & 0.2750 & 0.2651 & 0.2352 & 0.2599 & 0.4429 & {\bf 0.1035} \\ %gamma4
& (0.0361) & (0.0336) & (0.0363) & (0.0406) & (0.0393) & (0.0526) & (0.0396) & (0.0270) \\ \hline
Leukemia   & 0.0781 & 0.0763 & 0.1535 & {\bf 0.0456}& 0.2977 & 0.1617 & 0.0786 & 0.0525\\ %gamma2
& (0.0183) & (0.0227) & (0.0211) & (0.0244) & (0.0210) & (0.0786) & (0.0504) & (0.0704)\\ %gamma
\hline
\end{tabular}
\label{realdata}%
}
\end{center}
\end{table*}%

In Table \ref{realdata}, we observe that gSAVG generally improves on the performance of the average distance classifier and the scale adjusted version of it. This is expected in view of the facts that gSAVG resolves the issues with the existing classifiers in HDLSS scenarios. Among the competing methods, NN-RAND failed to yield promising results except {\tt GSE3726}. The GLMNET classifier is specifically designed for high-dimensional data with sparsity in their components, which is probably not the case in {\tt GSE3726} and {\tt HouseTwentys}. As a consequence, this classifier performed quite poorly except these two data sets. The RF classifier again lead to more comparable results (also see \citet*{FCB_2014}). 
SVM-RBF was far from being satisfactory, but interestingly, SVM-LIN showed slight edge over gSAVG for the data set {\tt Leukemia}.
\section{CONCLUDING REMARKS} \label{conclude}
In this article, we have proposed a new class of dissimilarity indices and modified the scale adjusted average distance classifier. Under appropriate conditions, we have proved that the misclassification probability for the resulting classifier goes to zero (i.e., {\it perfect classification}) in the HDLSS asymptotic regime. The methodology in this article is discussed for two-class classification problems. However, it can be easily extended for a multi-class problem. Using several simulated and real data sets, we have amply demonstrated improved performance of the proposed classifiers with respect to a wide variety of existing classifiers. We have considered three choices of $\gamma$ while analyzing the data sets. In practice, using a bounded $\gamma$ is advised, since the presence of outliers does not affect its performance.
\newpage

\bibliographystyle{apalike}
\bibliography{citation}
%\bibliography{citation1.tex}

\end{document}